\documentclass[12pt]{article}
\usepackage{verbatim}
\usepackage{amsfonts}
\usepackage{graphics}
\usepackage{amsmath}
\usepackage{times}
\usepackage{appendix}
\usepackage{color}
\usepackage{enumerate}
\usepackage{fancyhdr,latexsym,amsmath,amsfonts,amssymb,amsbsy,amsthm,url}
\usepackage[margin=0.5in,footskip=0.25in]{geometry}
\usepackage{graphics,graphicx,epsfig}
\usepackage{breqn}
\usepackage{caption,subcaption,float,subfloat}
\usepackage{pdflscape}
\usepackage{hyperref}

\usepackage[english]{babel}
\usepackage[dvipsnames]{xcolor}
\usepackage{tikz}

\numberwithin{equation}{section}

\fancypagestyle{usual}{
  \lhead[\thepage]{}
  \chead[{\it \shortauthors}]{\sc \shorttitle}
  \rhead[]{\thepage}
  \lfoot[]{}
  \cfoot[]{}
  \rfoot[]{}
  
}

\makeatletter

\renewcommand{\section}{
  \@startsection
  {section}
  {1}
  {0pt}
  {1.1\baselineskip}
  {0.2\baselineskip}
  {\sc \centering}
}

\renewcommand{\subsection}{
  \@startsection
  {subsection}
  {1}
  {0pt}
  {1.1\baselineskip}
  {0.2\baselineskip}
  {\sc \centering}
}

\renewcommand{\subsubsection}{
  \@startsection
  {subsubsection}
  {1}
  {0pt}
  {1.1\baselineskip}
  {0.2\baselineskip}
  {\sc \centering}
}

\makeatother

\usepackage[flushleft]{threeparttable}
\usepackage{rotating,booktabs,multirow}
\usepackage{ colortbl}
\usepackage{makecell, cellspace, caption}

\begin{document}

\title{\large\sc Did the lockdown curb the spread of COVID-19 infection rate in India: A data-driven analysis}
\normalsize
\author{\sc{Dipankar Mondal} \thanks{Department of Mathematics,
Indian Institute of Technology Guwahati, Guwahati-781039, Assam, India, e-mail: m.dipankar@iitg.ac.in}
\and \sc{Siddhartha P. Chakrabarty} \thanks{Department of Mathematics,
Indian Institute of Technology Guwahati, Guwahati-781039, Assam, India, e-mail: pratim@iitg.ac.in,
Phone: +91-361-2582606, Fax: +91-361-2582649}}
\date{}
\maketitle
\begin{abstract}

In order to analyze the effectiveness of three successive nationwide lockdown enforced in India, we present a data-driven analysis of four key parameters,
reducing the transmission rate, restraining the growth rate, flattening the epidemic curve and improving the health care system. These were
quantified by the consideration of four different metrics, namely, reproduction rate, growth rate, doubling time and death to recovery ratio.
The incidence data of the COVID-19 (during the period of 2nd March 2020 to 31st May 2020) outbreak in India was analyzed for the best fit to the epidemic curve, making use of the exponential growth, the maximum likelihood estimation, sequential Bayesian method and estimation of time-dependent reproduction. The best fit (based on the data considered) was for the time-dependent approach. Accordingly, this approach was used to assess the impact on the effective reproduction rate. The period of pre-lockdown to the end of lockdown 3, saw a $45\%$ reduction in the rate of effective reproduction rate. During the same period the growth rate reduced from $393\%$ during the pre-lockdown to $33\%$ after lockdown 3, accompanied by the average doubling time increasing form $4$-$6$ days to $12$-$14$ days. Finally, the death-to-recovery ratio dropped from $0.28$ (pre-lockdown) to $0.08$ after lockdown 3. In conclusion, all the four metrics considered to assess the effectiveness of the lockdown, exhibited significant favourable changes, from the pre-lockdown period to the end of lockdown 3. Analysis of the data in the post-lockdown period with these metrics will provide greater clarity with regards to the extent of the success of the lockdown.



\end{abstract}

\section{Introduction}
\label{Sec_Introduction}

As of 5th June 2020, the coronavirus disease 2019 (COVID-19) with its epicenter in Wuhan, China \cite{Kucharski20},
has resulted in more than $6.5$ million confirmed cases and $3,87,155$ causalities \cite{WHO20}. The global pandemic resulting from
COVID-19 was preceded by two other outbreaks of human coronavirus, in the 21st century itself, namely,
severe acute respiratory syndrome coronavirus (SARS-CoV) and Middle East respiratory syndrome coronavirus (MERS-CoV)
infections \cite{Gralinski20}. The possibilities of the source of the transmission of COVID-19 outbreak includes (but is not limited to)
animals, human-to-human and intermediate animal-vectors \cite{Gralinski20}. The index case for COVID-19 outbreak in India was reported
on 30th January 2020, in case of an individual with a travel history from Wuhan, China \cite{COVID19India}.
The data available on \cite{COVID19India}, suggests that during the early stages, the COVID-19 positive cases in India, were limited
to individuals with a travel history involving the global hotspots of the outbreak. However, subsequently, cases were detected in individuals who
neither had a travel history involving the global hotspots, nor had any contact with individuals who were already infected, which indicated the possibility of
community outbreak. This resulted in the Government of India announcing a lockdown across the country, driven by the necessity of ensuring that the social distancing norms are strictly observed. While the lockdown was not the only response to the pandemic, it was a very crucial step towards curbing the growth of COVID-19 in densely populated countries, like India. Given the concurrent economic cost of the lockdown, it is even more critical from the epidemiological
as well as economic perspective, to assess its effectiveness. This paper presents a data-driven analysis to examine the effectiveness of the lockdown, with an emphasis on the question as to whether the lockdown succeeded in curbing the intensity of COVID-19 spread rate in India ? In order to answer this, we
empirically analyze four different metrics, namely, reproduction number, growth rate, doubling time and death to recovery ratio, which quantify the
transmission rate, the growth rate, the curvature of epidemic curve and the improvement of health care capacity, respectively.

We now give a brief summary of some of the available literature on quantitative approaches to the modeling of transmission
of COVID-19 outbreak. A system of ordinary differential equation (ODE) driven model for phasic transmission of COVID-19, was analyzed for
calculating the transmissibility of the virus, in \cite{Chen20}. Kucharski et al. \cite{Kucharski20} considered a
stochastic transmission model on the data for cases in Wuhan, China (including cases that originated there) to estimate the likelihood
of the outbreak taking place in other geographical locations. A literature survey by Liu et al. \cite{Liu20}, summarized that
the reproductive number (and hence the infectivity) in case of COVID-19, exceeded that of SARS.
A Monte-Carlo simulation approach to assess the impact of the COVID-19 pandemic in India, was carried out in  \cite{Chatterjee20}.
In carrying out the mathematical and statistical modeling of COVID-19, it would be helpful to refer to the quantitative models
analyzed in case of the two preceding outbreaks of human coronavirus, namely SARS and MERS. In \cite{Small06},
a network model was analyzed to identify localized hotbeds, as well as super-spreaders for SARS.
Constrained by somewhat limited availability of data, a simple compartment model was used in \cite{Wang04}, for in-silico predictive
analysis of SARS outbreak in Beijing, China. Yan and Zou \cite{Yan08}, determined the optimal and sub-optimal strategies for quarantine
and isolation in case of SARS. A predictive model in \cite{Ejima14}, on imported cases of MERS, was used to ascertain the likelihood of a MERS
diagnosis, during the time window between immigration and onset of the disease. The trajectory of MERS outbreak was calibrated
to a dynamic model in \cite{Lee16}, with the goal of studying the role of time, in implementing the control measures.

A key identifier for the transmissibility of epidemiological diseases such as COVID-19 is the basic reproduction number
$R_{0}$, which is defined as the average number of secondary infections resulting from an infected case, in a population
whose all members are susceptible. Accordingly, we seek to estimate the data-driven value of $R_{0}$, for the outbreak of COVID-19 in India.
Further, we also seek to determine the time-dependent reproduction number $R_{t}$, for better clarity on the time-variability of the
reproduction number, particularly in the paradigm of its dynamics during the phases of the nationwide lockdown in India.
In addition, we also estimate and analyze the statistical performance of growth rate, doubling time and death to recovery ratio.

The paper is organized as follows. In Section \ref{Sec_Methodology},
we detail the source of the data as well as the statistical approaches used for the estimation of $R_0$ and $R_t$.
This will be followed by the discussion of the results for the outbreak in India, in Section \ref{Sec_Results}.
In Section \ref{Sec_Impact}, we present the data driven analysis of the impact of the lockdown.
And finally, in the concluding remarks in Section \ref{Sec_Conclusion}, we highlight the main takeaways for this analysis.

\section{Methodology for estimating reproduction rate}
\label{Sec_Methodology}

The data of incidences used for the analysis reported in this paper was obtained from the website of India COVID 19 Tracker \cite{COVID19India},
and used for the purpose of estimation of $R_{0}$. This estimation was carried out making use of the \texttt{R0} package \cite{Obadia12}
of the statistical package \texttt{R}. The standardized approach included in the \texttt{R0} package includes the implementation of the
Exponential Growth (EG), Maximum Likelihood (ML) estimation, Sequential Bayesian (SB) method and estimation of time dependent reproduction (TD)
numbers, used during the H1N1 pandemic of 2009. The package is designed for the estimation of both the ``initial'' reproduction number, as well as the ``time-dependent'' reproduction number. Accordingly, we present a brief summary of the four approaches used in the paper.

\begin{enumerate}

\item \textit{Exponential Growth (EG):}
As observed in \cite{Wallinga07}, the reproduction number can be indirectly estimated from the rate of the exponential
growth. In order to address the disparity in the different differential equation models, the authors observe that this disparity
can be attributed to the assumptions made about the shape of the generation interval distribution. Accordingly, the choice of the model,
used for the estimation of the reproduction number, is driven by the shape of the generation interval distribution.
Based on the assumption that the mean is equal to he generation intervals, the authors obtain the important result of determining
an upper bound on the possible range of values of the reproduction number for an observed rate of exponential growth, which manifests
into the worst case scenario for the reproductive number. Let the function $g(a)$ be representative of the generation interval distribution.
If the moment generating function $M(z)$ of $g(a)$ is given by $\displaystyle{M(z)=\int\limits_{0}^{\infty}e^{za}g(a)da}$, then the
reproduction number is given by $\displaystyle{R=\frac{1}{M(-r)}}$ subject to the condition that $\displaystyle{\frac{1}{M(-r)}}$
exists. In particular, the Poisson distribution can be used in the analysis of the integer valued incidence data \cite{Boelle09,Hens11},
for (discretized) generation time distribution. An important caveat is that this approach is applicable to the time window in which
the incidence data is observed to be exponential \cite{Obadia12}.

\item \textit{Maximum Likelihood (ML) estimation:}
The maximum likelihood model as proposed in \cite{White08} is based on the availability of incidence data $N_{0}, N_{1}, \dots, N_{T}$,
with the notation $N_{t},~t=0,1,2,\dots,T$ denoting the count of new cases at time $t$. In practice, we take the index $t$ in days, while noting
that this indexing is applicable for other lengths of time intervals. This approach is driven by the assumption that the Poisson distribution,
models the number of secondary infections from an index case, with the average providing the estimate for the basic reproduction number.
If we denote the number of observed incidences for consecutive time intervals by $n_{1}, n_{2}, \dots, n_{T}$ and
let $p_{i}$ denote the probability of the serial interval of a case in $i$ days (which can be estimated apriori), then the likelihood function is the
thinned Poisson: $\displaystyle{L\left(R_{0},\mathbf{p}\right)=\prod\limits_{t=1}^{T}\frac{e^{-\mu_{t}}\mu_{t}^{n_{t}}}{n_{t}!}}$. Note that here
$\displaystyle{\mu_{t}=R_{0}\sum\limits_{i=1}^{\min\left(k,t\right)}n_{t-i}p_{i}}$ and $\displaystyle{\mathbf{p}=\left(p_{1}, p_{2}, \dots, p_{k}\right)}$.
The absence of data from the index case can lead to an overestimation of the initial reproduction number, and accordingly a correction needs to be
implemented \cite{Obadia12}.

\item \textit{Sequential Bayesian (SB) method:}
A SIR model driven sequential estimation of the initial reproduction number was carried out by the sequential Bayesian method in \cite{Bettencourt08}.
It is based on the Poisson distribution driven estimate of incidence $n_{t+1}$ at time $t+1$ with the mean of $n_{t}e^{\gamma\left(R-1\right)}$.
In particular, the probability distribution for the reproduction number $R$, based on the observed temporal data is given by
$\displaystyle{P\left[R|n_{1},n_{1},\dots,n_{t+1}\right]=\frac{P\left[n_{1},n_{1},\dots,n_{t+1}|R\right]P\left[R\right]}
{P\left[n_{1},n_{1},\dots,n_{t+1}|R\right]}}$, where $P\left[R\right]$ is the prior distribution of $R$ and
$P\left[n_{1},n_{1},\dots,n_{t+1}\right]$ is independent of $R$.

\item \textit{Estimation of time dependent reproduction (TD):}
The TD method is amenable to the computation of the reproduction numbers through the averaging over all networks of transmission, based on the
observed data \cite{Wallinga04}. Let $i$ and $j$ be two cases, with the respective times of onset of symptoms being $t_{i}$ and $t_{j}$.
Further, let $p_{ij}$ denote the probability of $i$ being infected by $j$. If $g(a)$ denotes the distribution of the generation interval,
then $\displaystyle{p_{ij}=\frac{g\left(t_{i}-t_{j}\right)}{\sum\limits_{i\ne k}w\left(t_{i}-t_{k}\right)}}$. Accordingly, the
effective reproduction number is given by $\displaystyle{R_{j}=\sum\limits_{i}p_{ij}}$, whose average is then given by
$\displaystyle{R_{t}=\frac{1}{n_{t}}\sum\limits_{t_{j}=t}R_{j}}$. In absence of observed secondary cases, a correction can be made to the
time dependent estimation \cite{Cauchemez06}

\end{enumerate}

\section{Estimating the reproduction numbers and fitting the epidemic curve}
\label{Sec_Results}

\begin{figure}[H]
\centering
\includegraphics[width=0.60\textwidth]{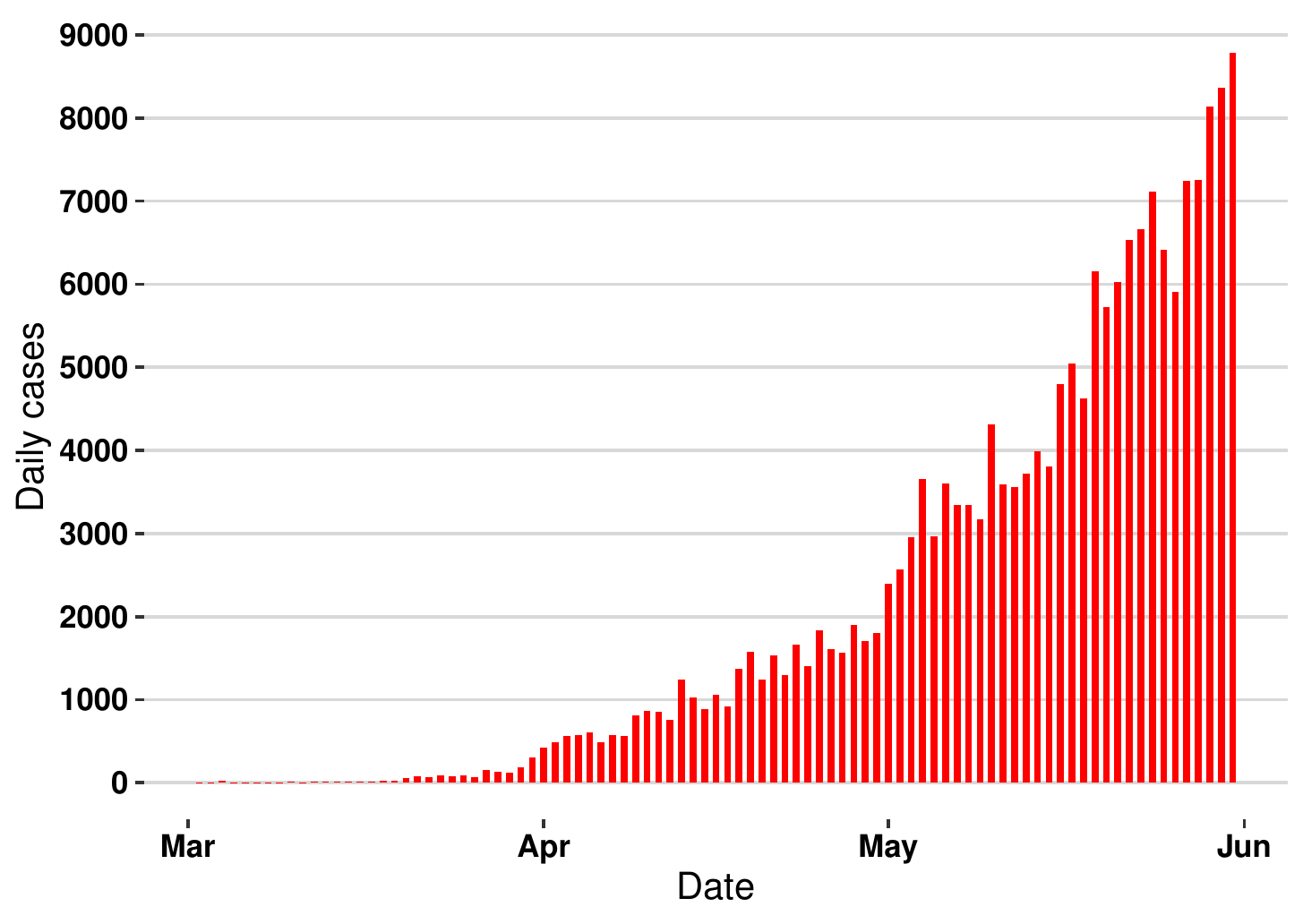}
\caption{Epidemic curve for the period of 2nd March 2020 to 31st May 2020}
\label{dly}
\end{figure}

In this section, we undertake the fitting of the epidemic curve and the estimation of the reproduction numbers using the approaches
enumerated in Section \ref{Sec_Methodology}.
We have obtained the daily incidence data, for the period of 2nd March 2020 to 31st May 2020 \cite{COVID19India}.
The epidemic curve based on the  data, for this period, is depicted in Figure \ref{dly}, which indicates that the number of COVID-19 positive cases,
were growing in an almost exponential manner.

The initial reproduction number $R_{0}$ according to the EG is $1.339$, with the $95\%$ confidence interval (CI) for this estimation being
$[1.337, 1.341]$. For the case of the ML method, the $R_{0}$ is determine to be $1.26$ and the corresponding $95\%$ CI is $[1.257, 1.273]$.

\begin{table}[H]
\centering	
\begin{tabular}{|c|c|c|}
\bottomrule
Method & $R_0 / R_{t}$  & 95\%-CI   \\
\hline
EG &  1.339 & $[1.337, 1.341]$  \\
\hline
ML & 1.26 & $[1.257, 1.273]$ \\
\hline
SB & 1.591 & $[1.285, 1.984,]$ \\
\hline
TD & 1.68 & $[1.41, 2.012]$ \\
\toprule
\end{tabular}
\caption{Initial and time varying reproduction numbers using the four methods}
\label{t1}
\end{table}

\begin{figure}[H]
\centering
\includegraphics[width=0.60\textwidth]{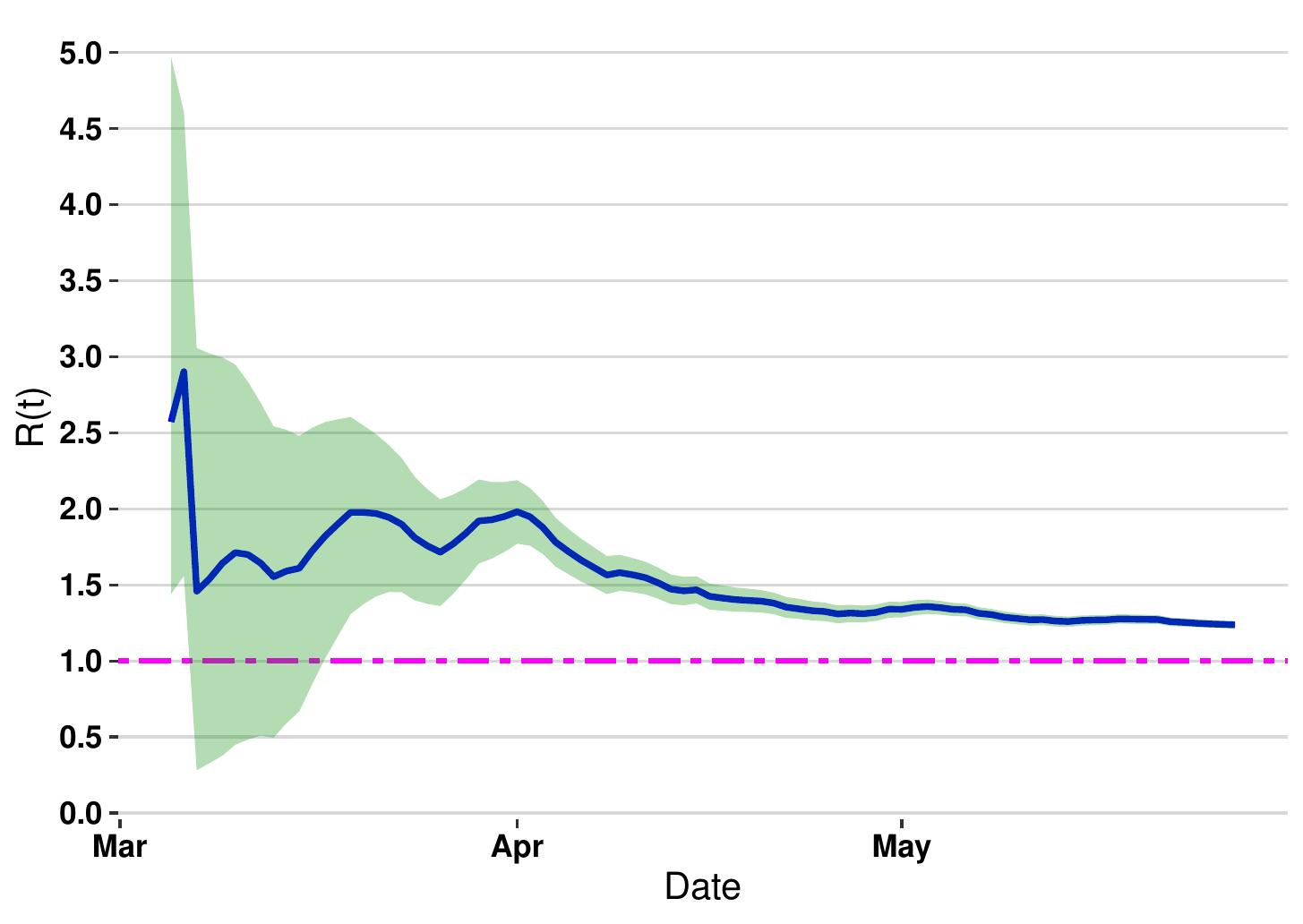}
\caption{Seven-day rolling $R_{t}$ using SB Method}
\label{fig:Rp-SB}
\end{figure}

\begin{figure}[H]
\centering
\includegraphics[width=0.60\textwidth]{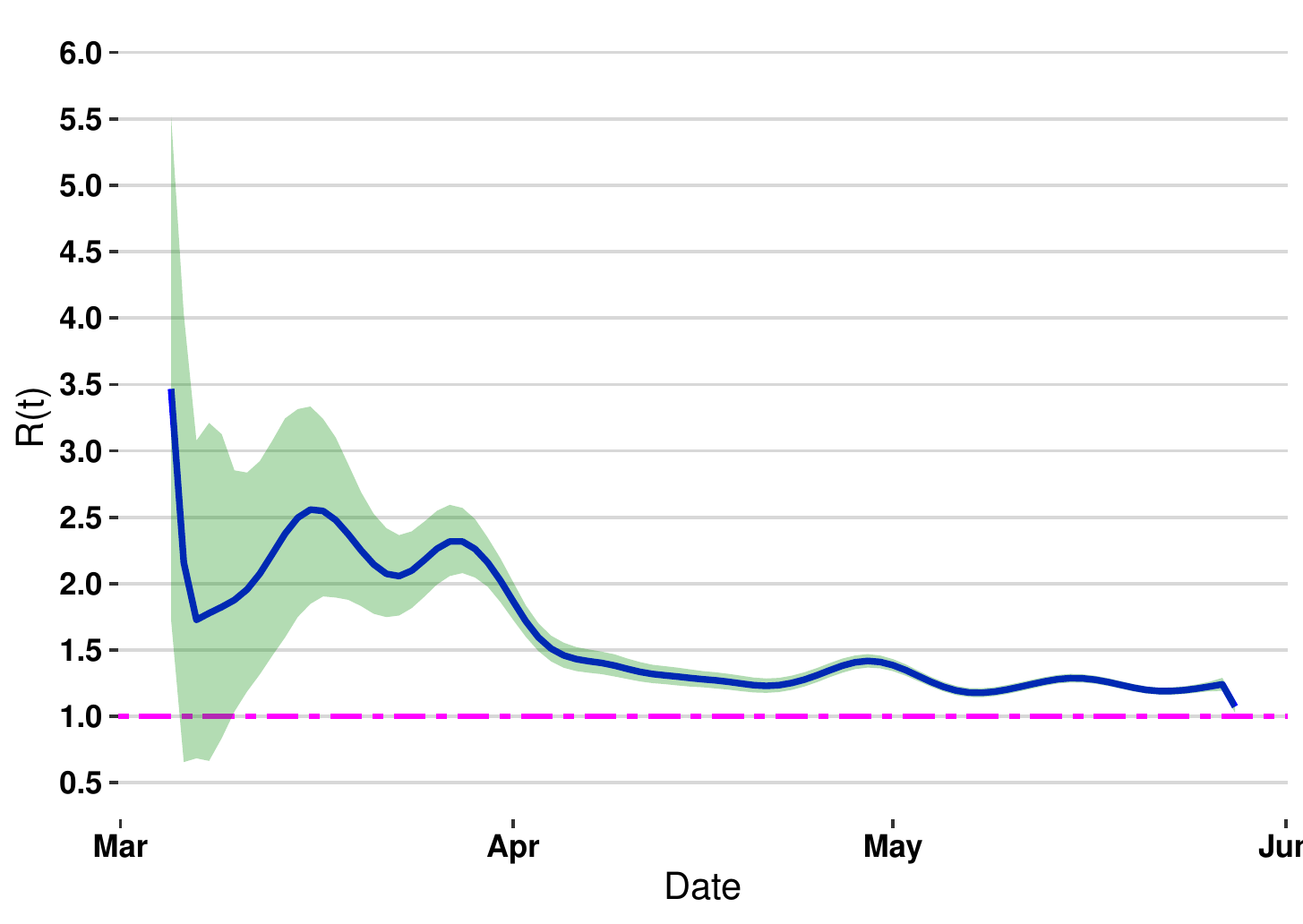}
\caption{Seven-day rolling $R_{t}$ using TD method}
\label{fig:tRp-TD}
\end{figure}

For the estimation of time-varying  reproduction numbers or the effective reproduction numbers $R_{t}$, the generation time distribution is required.
Accordingly, we use  gamma distribution with mean of $5.2$ days and the standard deviation of $2.8$ days as reported from China \cite{Ganyani20}.
Now, the average $R_{t}$, using the SB and TD methods, are $1.591$ and $1.68$, respectively. The $R_{0}$ values using EG and ML, and the
$R_{t}$ values using SB and TD, along with the corresponding $95\%$ confidence intervals are tabulated in Table \ref{t1}.
Further, the seven-day rolling $R_{t}$, obtained for the cases of SB and TD, are plotted in Figure \ref{fig:Rp-SB} and Figure \ref{fig:tRp-TD}, respectively.


\begin{figure}[H]
\centering
\begin{subfigure}[b]{0.40\textwidth}
\includegraphics[width=\textwidth]{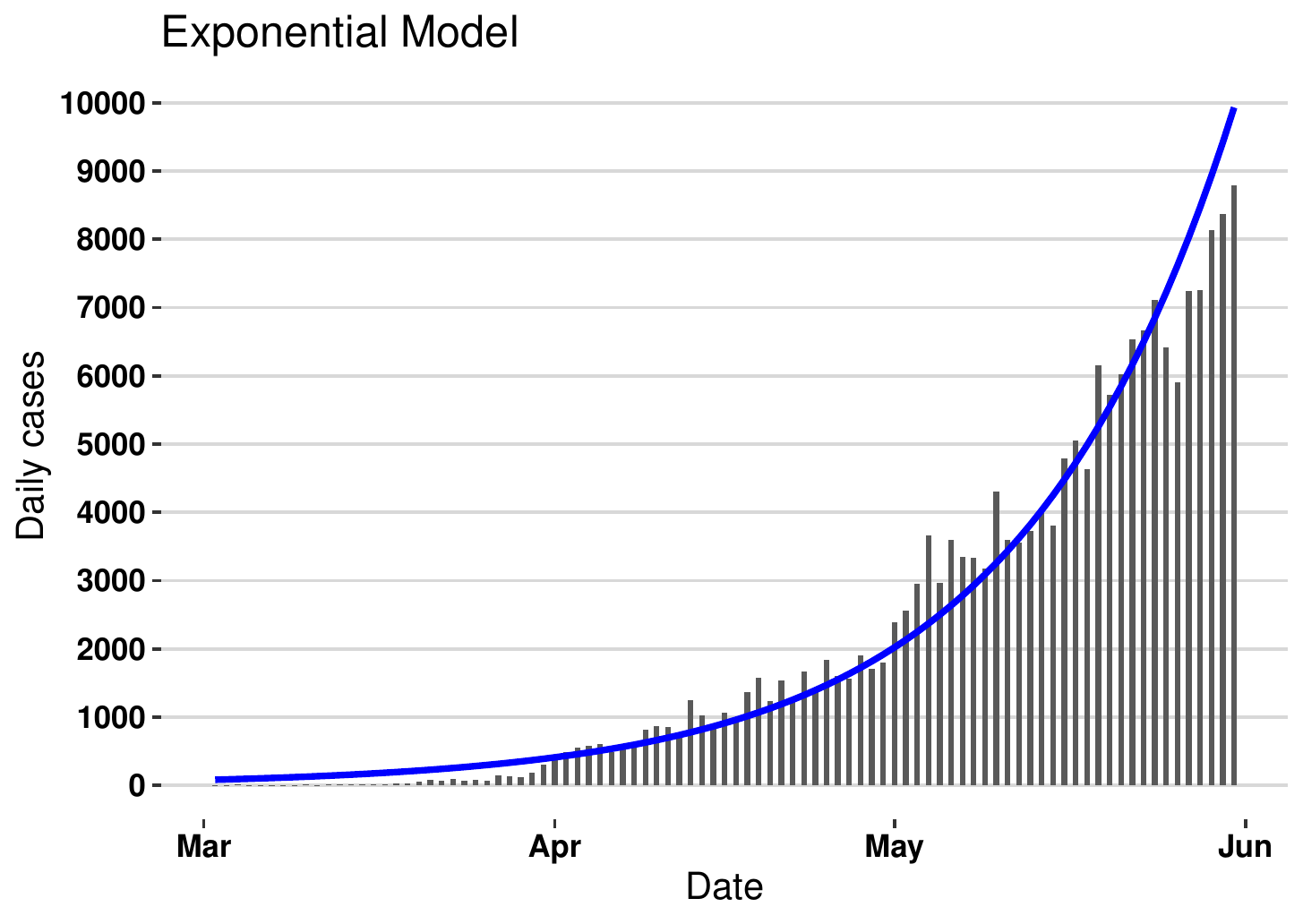}
\caption{EG}
\end{subfigure}
\hspace{2 cm}
\begin{subfigure}[b]{0.40\textwidth}
\includegraphics[width=\textwidth]{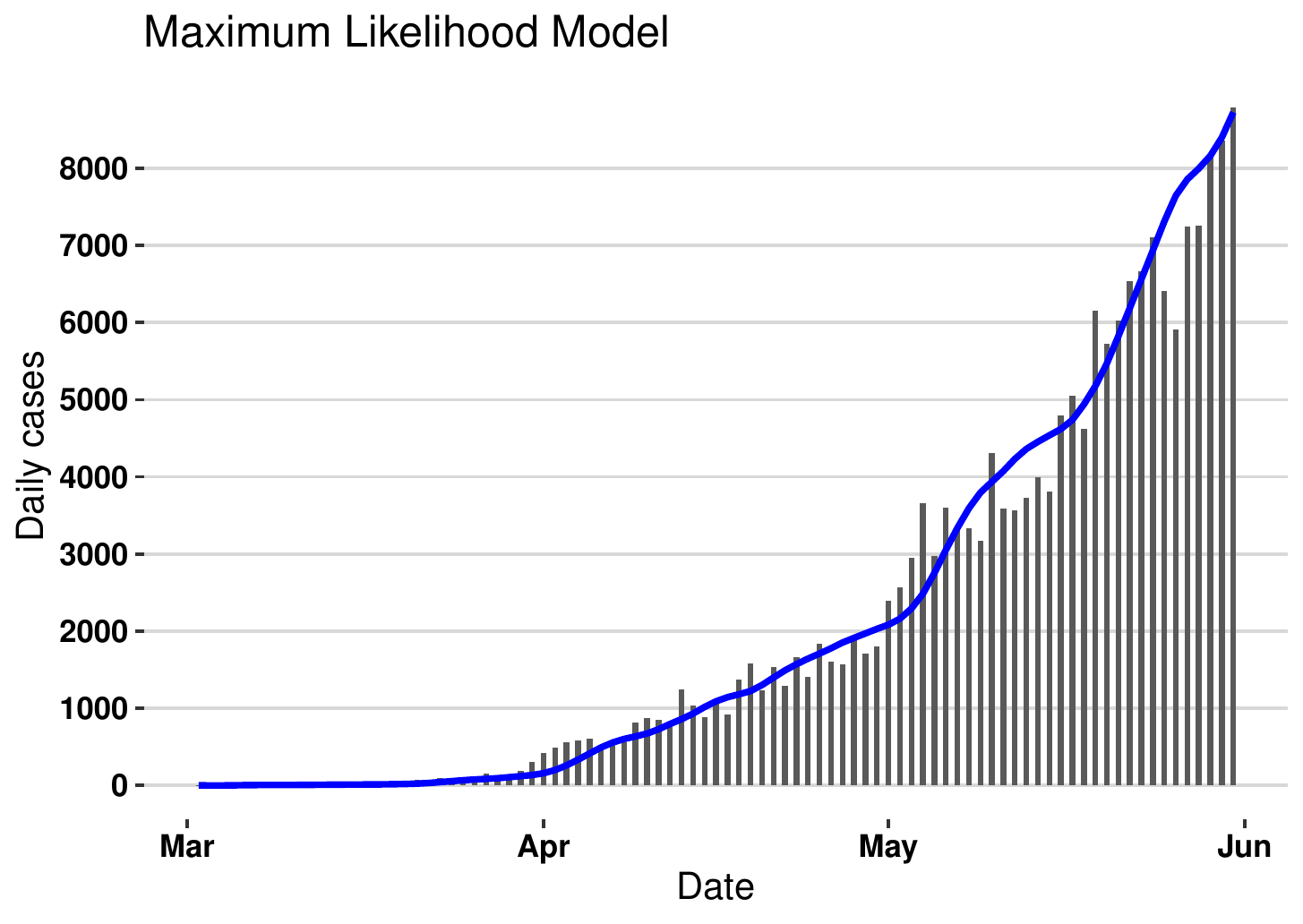}
\caption{ML}
\end{subfigure}
\begin{subfigure}[b]{0.40\textwidth}
\includegraphics[width=\textwidth]{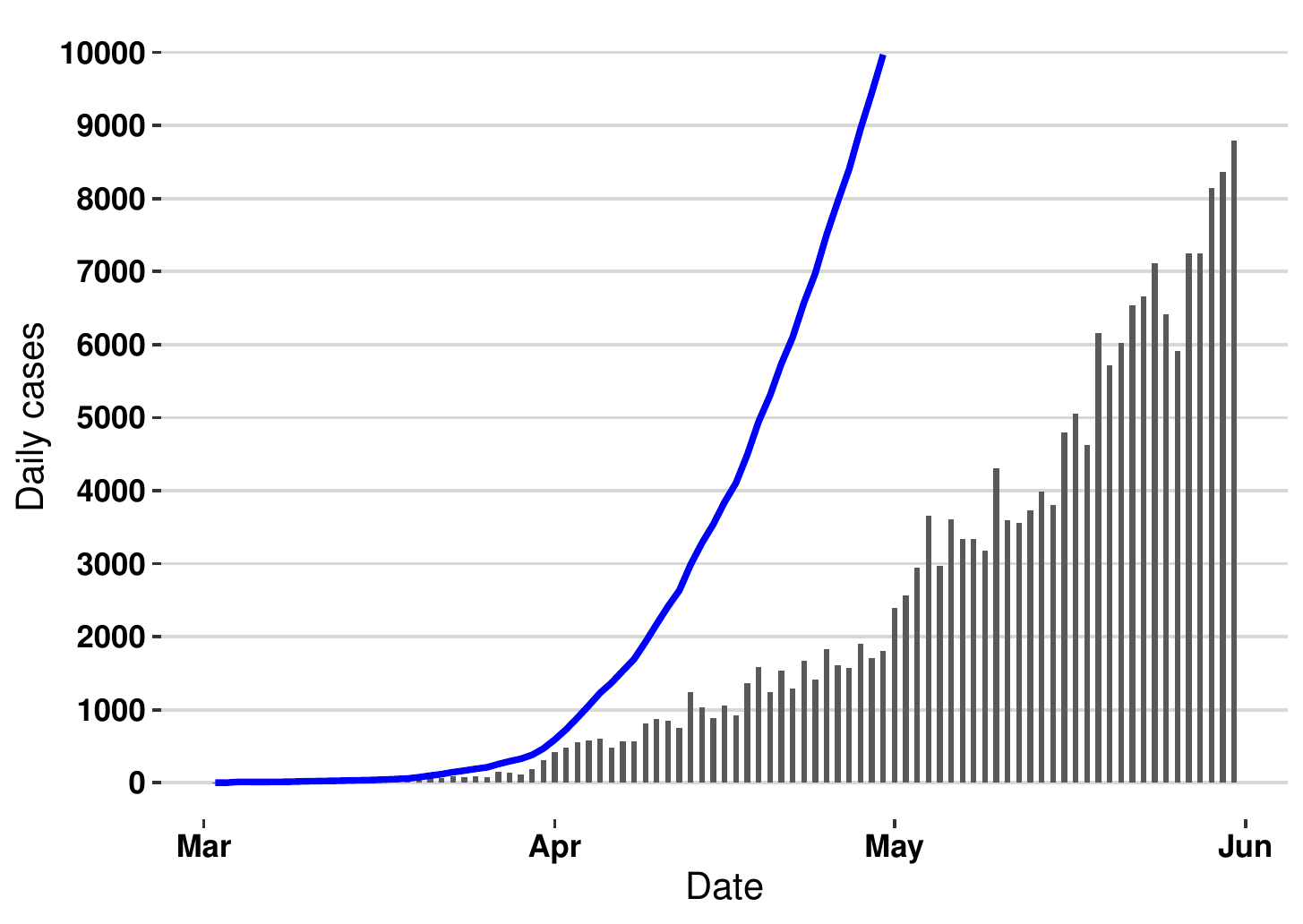}
\caption{SB}
\end{subfigure}
\hspace{2 cm}
\begin{subfigure}[b]{0.41\textwidth}
\includegraphics[width=\textwidth]{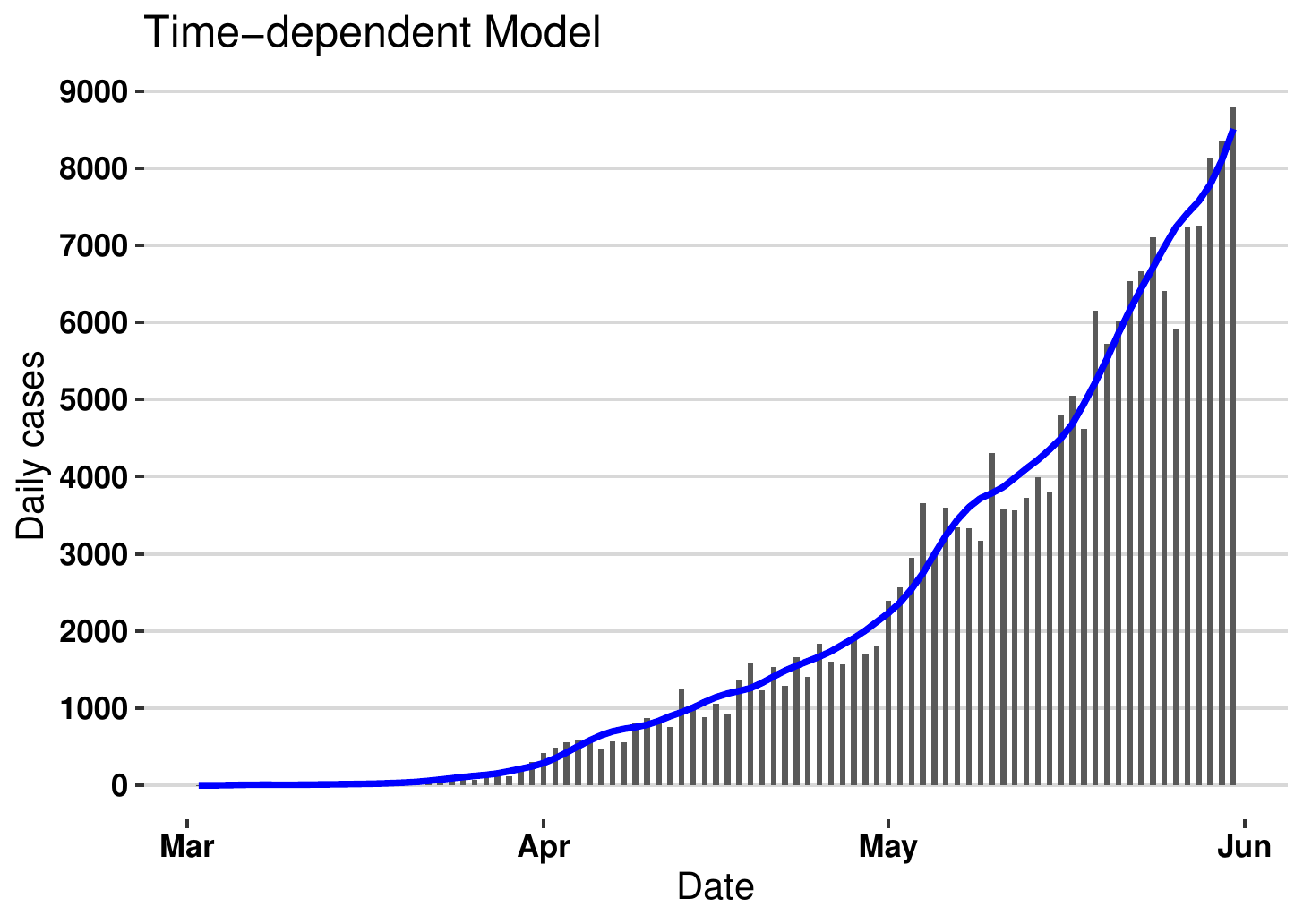}
\caption{TD}
\end{subfigure}
\caption{Epidemic curve using the four methods}
\label{fig:ec-four}
\end{figure}

Besides estimating the reproduction rate, we fit the epidemic curve, making use of the four models, namely EG, ML, SB and TD.
Accordingly, the predicted incidence (based on the fitted model parameters in each case) and the
observed incidence for each method, are illustrated in Figure \ref{fig:ec-four}.

The prediction provided by the EG, ML and TD, are reasonably close to the actual cases. However, it is clearly observed that the most poorly fitted
model is the SB model. The SB model overestimates the epidemic curve, and thus the predictions according to this model are much higher than the actual incidences. Therefore, in order to find the best-fitted model, the root mean squared errors,
$\displaystyle{RMSE:=\sqrt{\sum\limits_{i=1}^{n}\frac{(\hat{y_i}-y_i)^2}{n}}}$,
for all the models were calculated. As expected, the RMSE for the SB model is the highest. On the other hand, the TD model has the lowest RMSE.
The RMSE values for all the four models are tabulated in Table \ref{tab:RMSE}, from where we can conclude (based on the data set considered)
that the best model for the estimation of the COVID-19 epidemic in India, is the TD model.

\begin{table}[H]
\centering	
\begin{tabular}{|cccc|}
\bottomrule
EG & ML & TD & SB\\
\hline
430 & 368 & 289 & 15377 \\
\toprule
\end{tabular}
\caption{RMSE for the four methods}
\label{tab:RMSE}
\end{table}

\section{Impact of lockdown}
\label{Sec_Impact}

The nationwide lockdown was imposed, on 25th March, 2020, with the goal of arresting the spread of infection, through strict restrictions on
mass movement and encouraging social distancing, and it was expected that the spread rate would come down, along with the reduction in the
possibility of community transmission.
This in turn would result in curbing the number of cases from rising dramatically, thereby enabling the healthcare
system with more time to make necessary arrangements for the better preparedness of the medical infrastructure.
Thus, the first phase of lockdown until 14th April, 2020, was extended to another two phases of lockdown, with slightly relaxed restrictions,
and were enforced from 15th April to 3rd May, 2020 and from 4th May to 31st May, 2020.
This section discusses the impact of the entire lockdown on COVID-19 spread, by analyzing various metrics, namely,
the effective reproduction rate, the growth rate, the doubling time and the death to recovery ratio.

\subsection{Impact on effective reproduction rate}

One of the key mathematical indicator relied upon, in the paradigm of the spread of COVID-19 pandemic and consequent policy decisions is the
effective reproduction rate (ERR) or the time-varying reproduction number. As ERR provides the information of time varying transmission rate,
it would be a natural choice to measure the impact of the entire lockdown, as well as different phases of the lockdown.
In the preceding Section \ref{Sec_Results}, we have shown that, amongst all the models, the TD is the best fitted model, for the Indian epidemic curve.
Hence, we discuss the impact of lock-down in the context of the TD-based $R_{t}$.

\begin{figure}[H]
\centering
\begin{subfigure}[b]{0.40\textwidth}
\includegraphics[width=\textwidth]{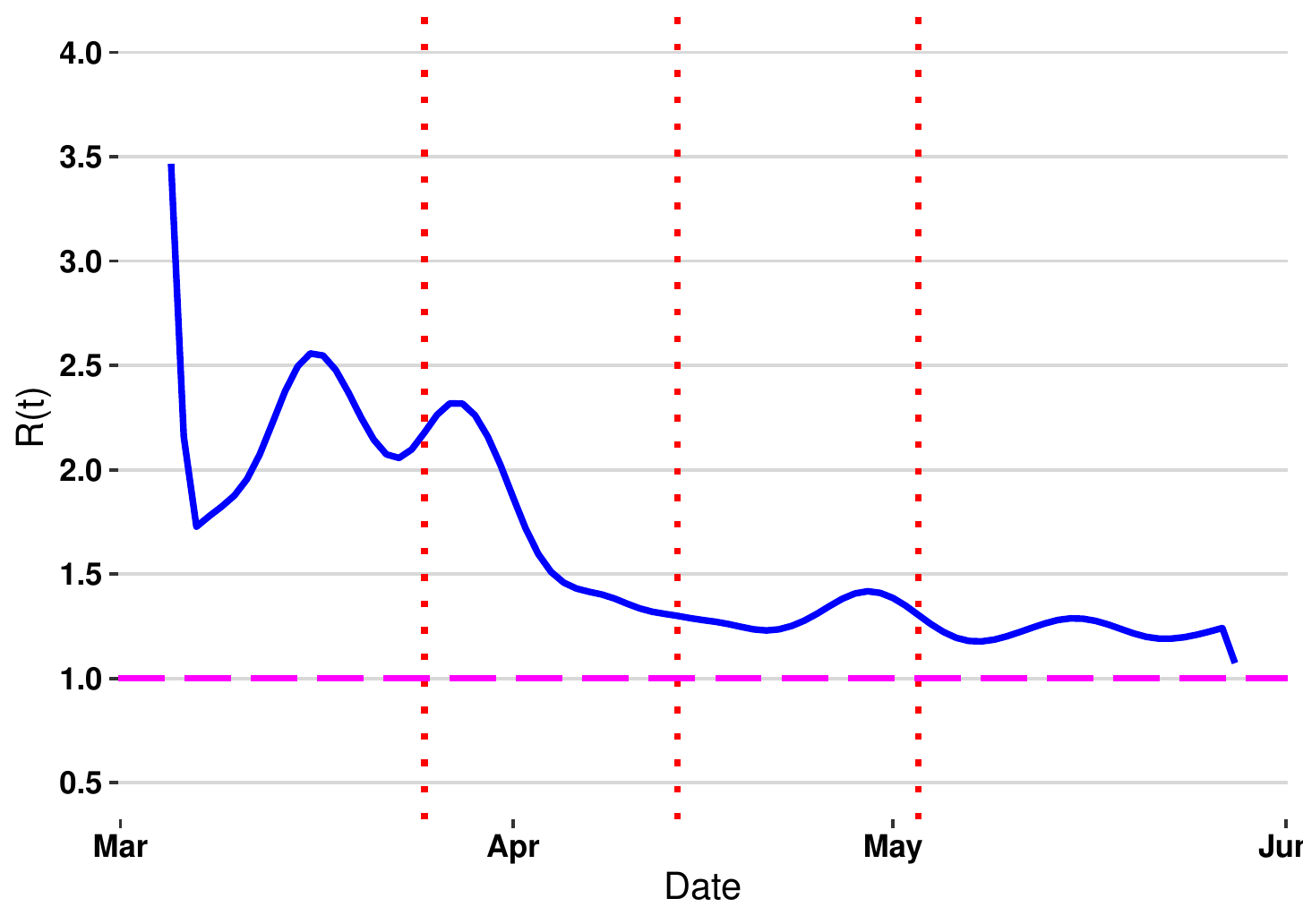}
\caption{$R_{t}$ using TD}
\label{r1td}
\end{subfigure}
\hspace{2 cm}
\begin{subfigure}[b]{0.40\textwidth}
\includegraphics[width=\textwidth]{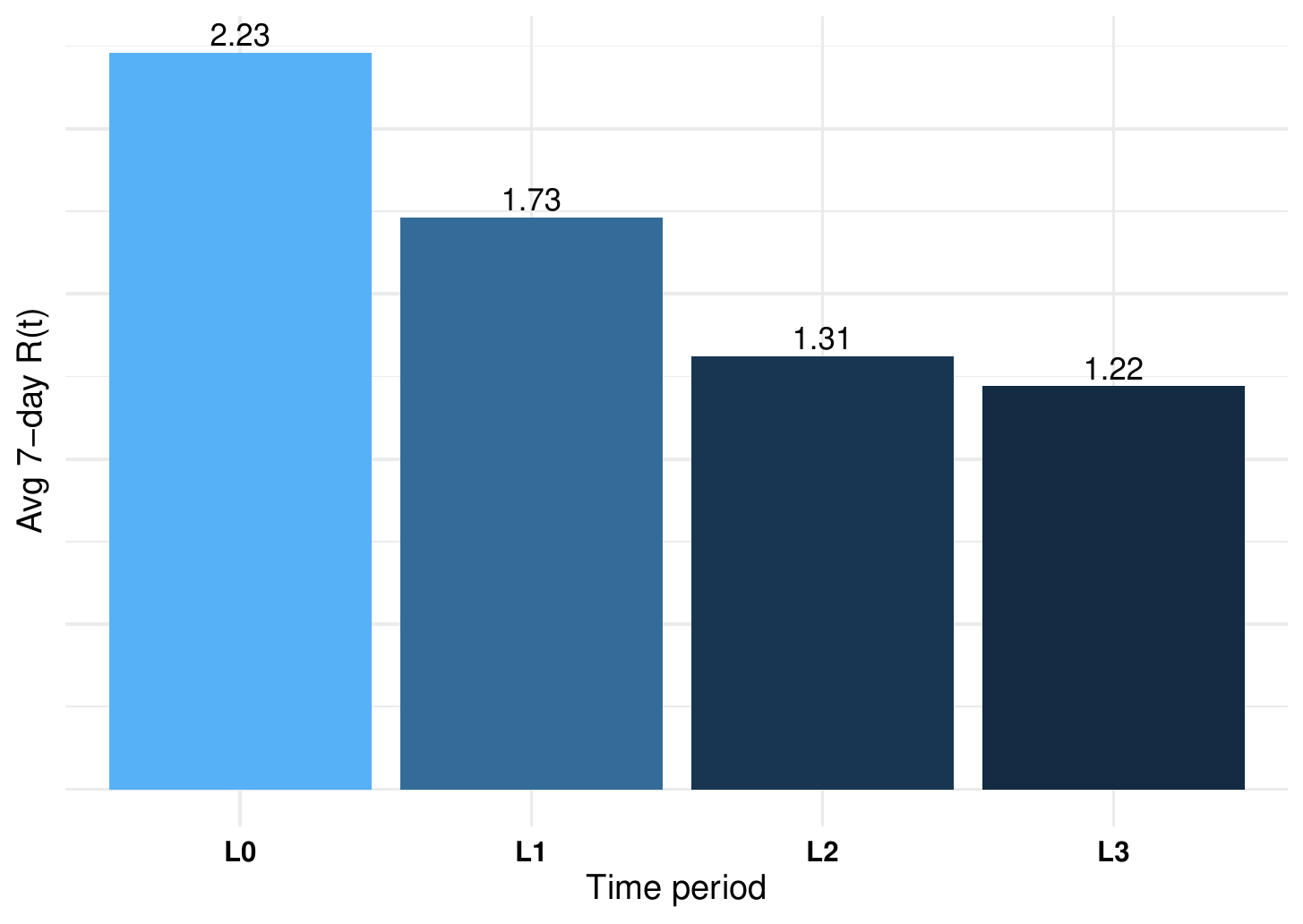}
\caption{Average seven-day $R_{t}$}
\label{f2b}
\end{subfigure}
\caption{Impact of lockdown on ERR}
\label{fig:epr-td}
\end{figure}

Figure \ref{r1td} depicts the seven-day rolling ERR. It is clearly observed that, before the lockdown, the $R_{t}$ was unsteady,
but it started dipping downward after the commencement of the lockdown. In the pre-lockdown period, the average seven-day ERR  was 2.23.
Therefore, before the lockdown, if $100$ individuals had COVID-19, they would have infected $223$ people on an average. In the first lock-down period,
the average ERR came down to $1.73$, a 22\% drop. Thus, at this rate, $100$ carriers would infect $173$ others on an average. In the second and third lockdown periods, the ERR furthers dipped to $1.31$ and $1.22$, respectively. Therefore, from the pre-lockdown to the end of lockdown 3, the overall rate of
reduction of ERR was nearly $45\%$. Figure \ref{f2b} displays the phase-wise \footnote{L0, L1, L2 and L3 imply  pre-lockdown, lockdown 1, lockdown 2
and lockdown 3, respectively} average $R_{t}$. The descriptive statistics of $R_{t}$ and the corresponding confidence intervals are
described in Table \ref{t32}. From these results, we can clearly infer that, so far, the lockdown has by and large succeeded, in reducing the ERR.
However, this observation come with the caveat that the three successive lockdowns did not drive the $R_{t}$ below $1$, which is
suggestive that the epidemic may exhibit a surge once all the restrictions are lifted.

\begin{table}[H]
\centering
\begin{tabular}{|l|ccc|c|}
\bottomrule
{\multirow{2}{*}{Periods}} & \multicolumn{3}{c|}{{7-day ERR}} &  	{\multirow{2}{*}{95\%-CI}} \\
\cline{2-4}
& Min & Max & Average & \\
\hline
L0 & 1.73 & 3.466 & 2.23  & $[1.468, 3.108]$ \\
L1 & 1.30 & 2.317 & 1.73  & $[1.592, 1.875]$ \\
L2 & 1.23 & 1.42  & 1.31  & $[1.256, 1.365]$ \\
L3 & 1.07 & 1.29  & 1.22  & $[1.188, 1.254]$ \\
\toprule
\end{tabular}
\caption{Reproduction rate and $95\%$-CI in different periods}
\label{t32}
\end{table}

\subsection{Impact on growth rate}

The reduction of ERR should further reduce the growth rate of daily incidences. In order to see the growth rate, in a particular time period,
we calculate the seven-day rolling growth rate in that period, and then take the average. Suppose that we have daily incidence numbers,
$D(t),~t=1,2,3,\dots,20,$ for a period of $20$ days. We first compute the seven-day rolling growth rates,
$\displaystyle{\frac{D(i+7)-D(i)}{D(i)}}$, where $i=1,2,3,\dots,13$, and we get a dataset of $13$ points. Finally, the simple mean of the dataset
is calculated. If the seven-day average growth is $30\%$ in a month, then  the average weekly number of positive cases would have increased from
$100$ to $130$ in that month.

\begin{figure}[H]
\centering
\includegraphics[width=0.60\textwidth]{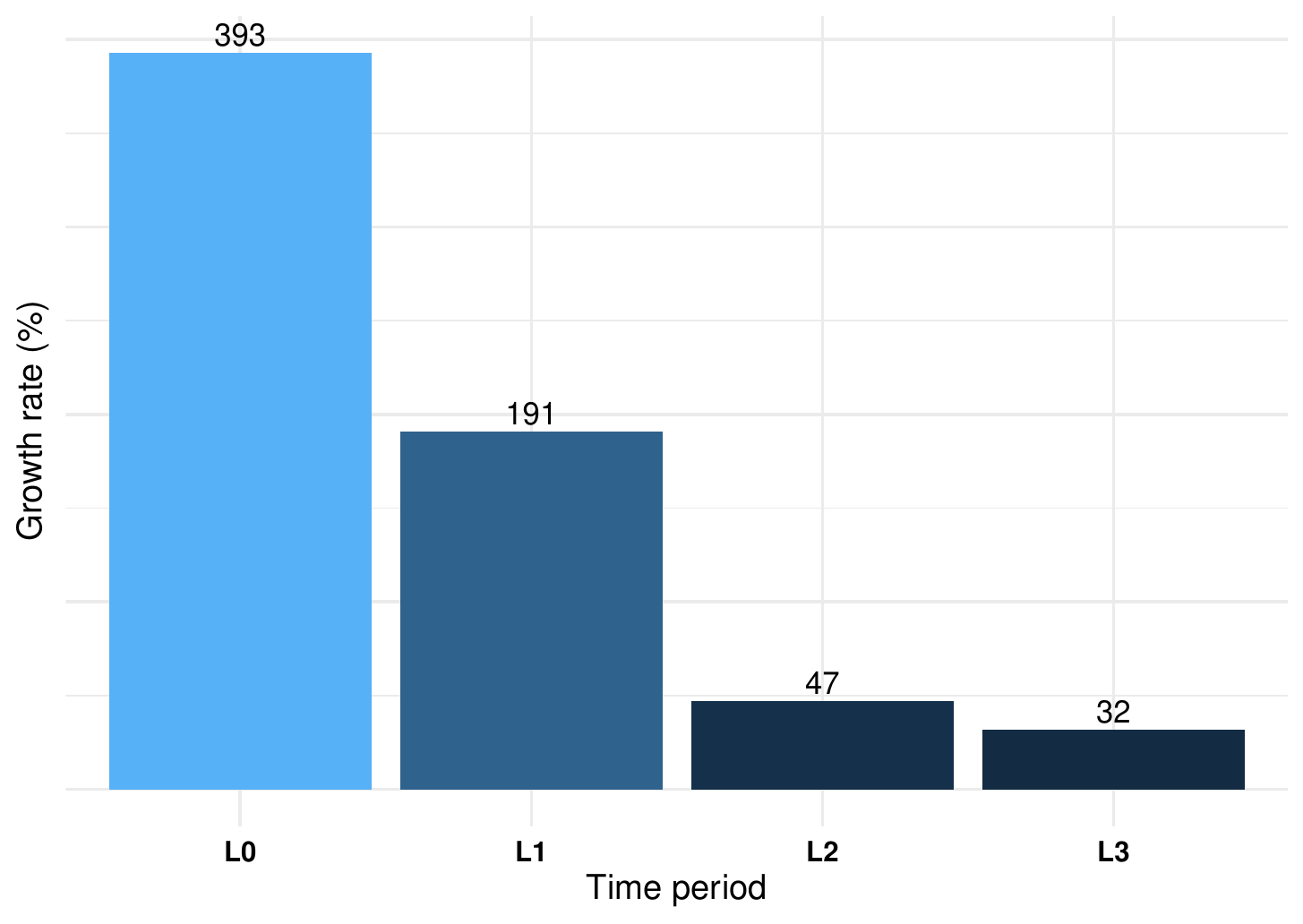}
\caption{Weekly growth rate of positive cases}
\label{gr}
\end{figure}

Figure \ref{gr} illustrates the average weekly growth rate in different time periods. In the pre-lockdown period (L0), the growth rate was 393\%.
It means that the weekly number of positive cases, increased drastically from $100$ to $493$ in the pre-lockdown period. The growth rate has  decreased
to $191\%$ in lockdown 1 (L1). It further reduced to $47\%$ and $32\%$ in lockdown 2 (L2) and lockdown 3 (L3), respectively.
Therefore, we can conclude, that the implementation of nationwide lockdown has resulted in slowing down the growth rate of COVID-19 positive cases.

\subsection{Impact on doubling time}

One of the key indicator to see the spread of any pandemic is the doubling time. It is referred to as the time (usually counted in number of days)
it takes for the total number of cases to double. The doubling time of $n$ days means that if there were $100$ cases at day 0, then, on day $n$,
the number of cases would be $200$. The more the doubling time is, the more the possibility of achieving a flattened epidemic curve.

\begin{figure}[H]
\centering
\begin{subfigure}[b]{0.40\textwidth}
\includegraphics[width=\textwidth]{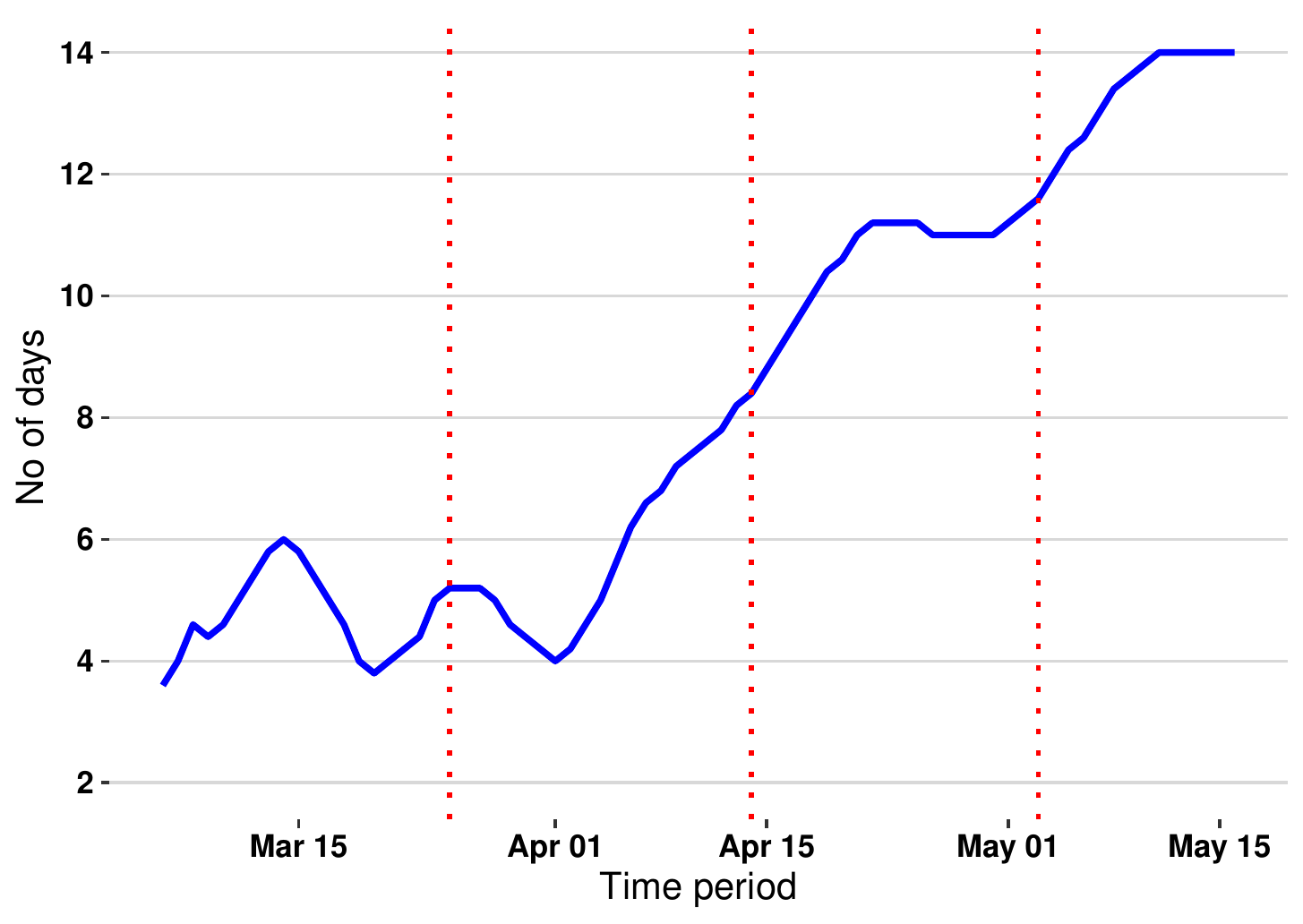}
\caption{Doubling time}
\label{dbr1}
\end{subfigure}
\hspace{2 cm}
\begin{subfigure}[b]{0.40\textwidth}
\includegraphics[width=\textwidth]{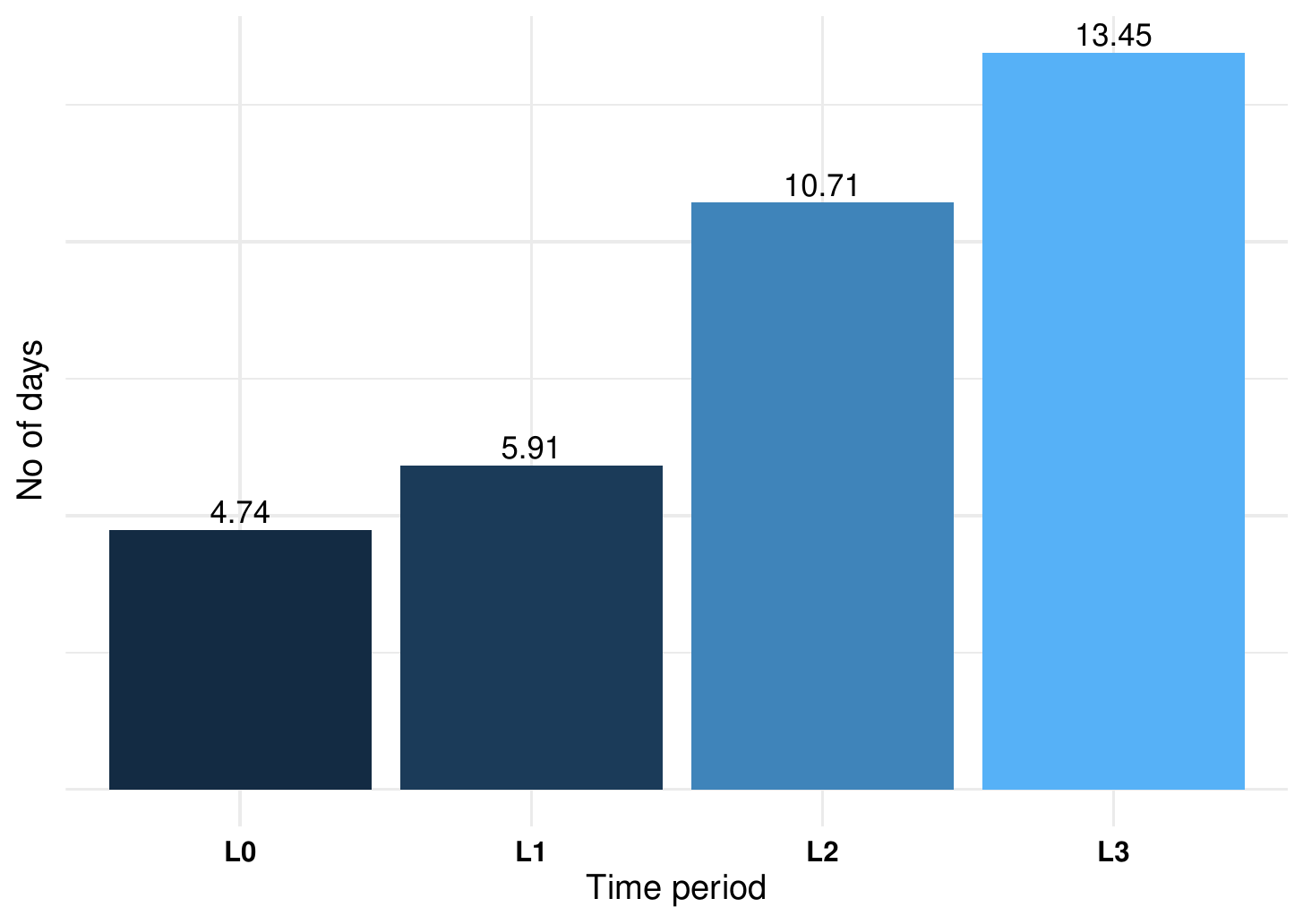}
\caption{Average doubling time in different periods}
\label{dbr2}
\end{subfigure}
\caption{Impact of lockdown on doubling time}
\label{fig:dt-td}
\end{figure}

Figure \ref{dbr1} displays the doubling rate for five-day moving averages. The escalation in doubling time is easily seen from the figure.
The doubling time during third lockdown period was about $12$-$14$ days, up from $4$-$6$ days prior to the commencement of the lockdown.
The phase-wise average doubling timings are shown in Figure \ref{dbr2}. The increment in doubling time is clearly visible from this figure.
Therefore, from these results, we infer that the doubling time has improved significantly after the enforcement of nationwide lockdown.

\subsection{Impact on death to recovery ratio}

\begin{figure}[H]
\centering
\begin{subfigure}[b]{0.40\textwidth}
\includegraphics[width=\textwidth]{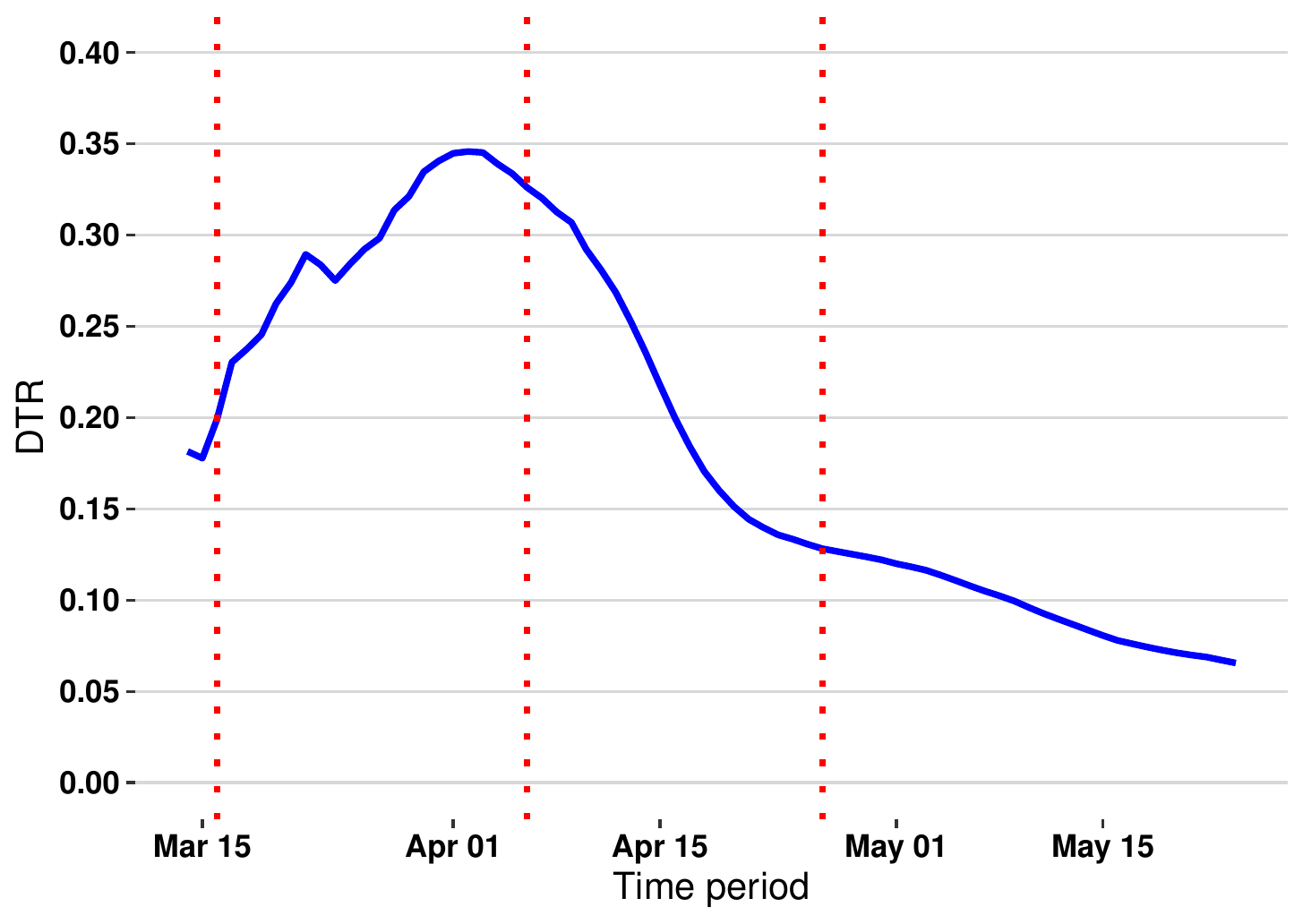}
\caption{Death to recovery ratio}
\label{drt1}
\end{subfigure}
\hspace{2 cm}
\begin{subfigure}[b]{0.40\textwidth}
\includegraphics[width=\textwidth]{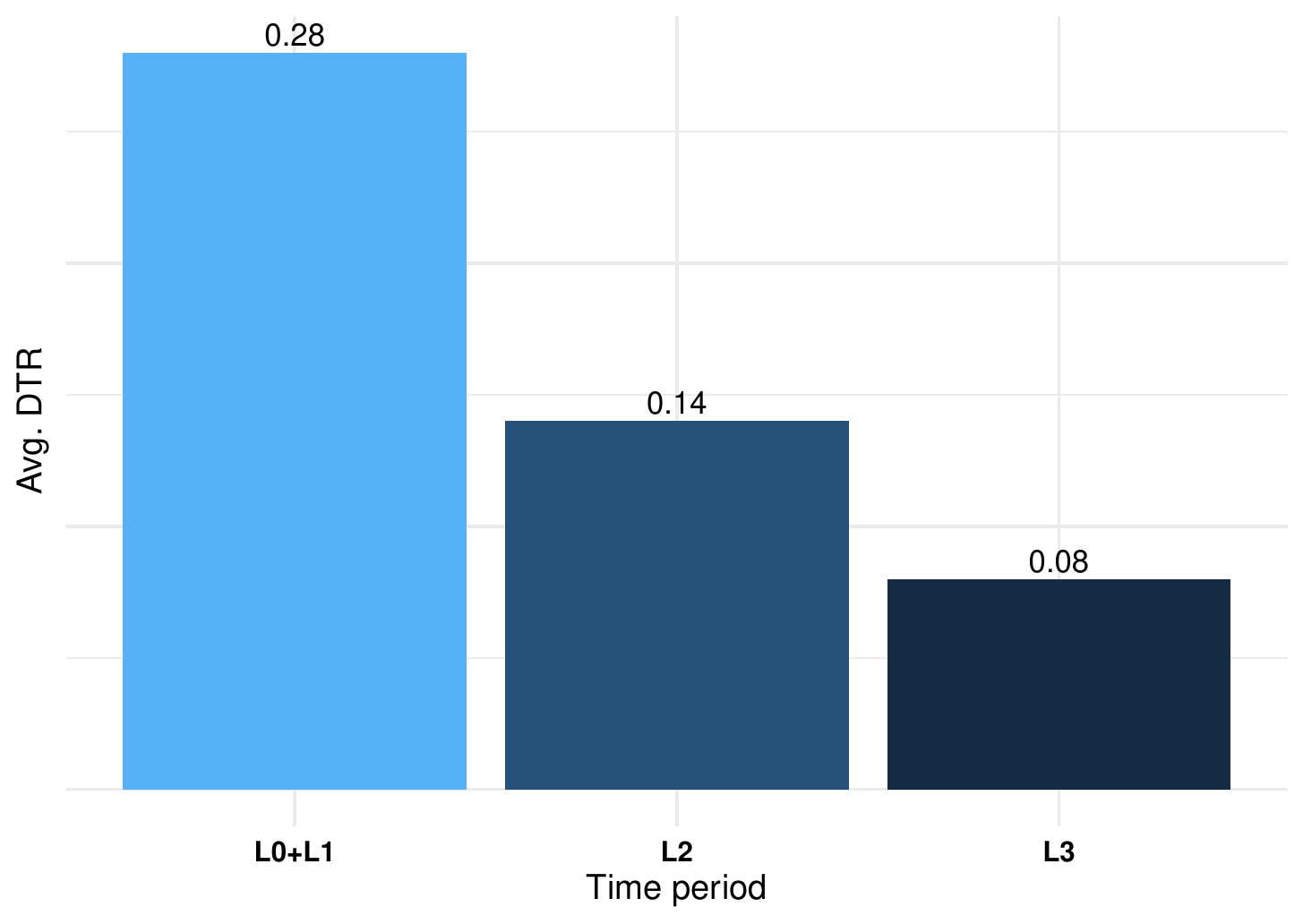}
\caption{Average death to recovery ratio}
\label{avdrt}
\end{subfigure}
\caption{Impact of lockdown on death to recovery ratio}
\label{fig:impact-dtr}
\end{figure}

In a pandemic, the performance of any nation's health care system, is measured ultimately in terms of deaths and recoveries.
This segment discuses the effect of lockdown on death to recovery ratio (DTR).
The DTR is defined as a ratio between total number of deaths and total number of recoveries:
\[DTR_{t}=\frac{\mbox{Total number of deaths upto time t}}{\mbox{Total number of recoveries upto time t}}.\]
The DTR stipulates the clinical management ability or the efficiency of health system. It is highly important to keep the value of the DTR
as low as possible. Mathematically, the closer this value is to zero, the better the efficiency of healthcare system, in dealing with the pandemic.
For example, $DTR_{t} = 0.5$ implies that, for every $100$ recoveries, $50$ infected patients would have died. The seven-day rolling DTR  is
plotted in Figure \ref{drt1}. It is clearly seen that the DTR has declined  significantly as time has progressed. The phase-wise bar chart also
depicts the reduction of DTR over the period of three months. In pre-lockdown (L0) and lockdown 1 (L1) periods together, the average DTR was 0.28.
It reduced to $0.14$ in lockdown 2 (L2) and further declined to $0.08$ in lockdown 3 (L3), which shows that, in this short period,
the Indian health care system has been improved significantly to tackle the COVID-19 pandemic.

\section{Conclusion}
\label{Sec_Conclusion}

In this paper, we have discussed the impact of lockdown on COVID-19 infection rate, in India. The aim was to see whether the lockdown has really
curbed intensity of spread. In order to do that, we empirically analyzed different metrics that mainly measure the spread of infectious disease, like COVID-19.
The metrics are effective reproduction rate, growth rate, doubling time and death to recovery ratio (DTR).
For case of ERR, it is seen that the lockdown has reduced the reproduction rate by more than $40\%$. The growth rate has also
substantially decreased from the initial period to the end of lockdown. On the other hand, the doubling time has largely improved over the
three month period. The rate of increment from pre-lockdown to lockdown 3 is nearly $183\%$. Finally, we described the impact on DTR, which quantifies
the number of death against the number of recoveries. We observed significant downfall of DTR from the month of April. On average, the initial DTR of $0.28$
has dipped downward to $0.08$ at the third phase of lockdown.  Therefore, despite rising cases of COVID-19 infection in India, the lockdown has managed to
curb the spread to  some extent. However, the caveat is that, despite the encouraging results, the pandemic will persist, unless the ERR is driven below
$1$. It remains to be seen if there is a adverse movement of the metrics, after the relaxation of the restrictions.
The behaviour of these metrics in post-lockdown period will provide a more accurate and complete information regarding the success or failure of lockdown.

\section*{Acknowledgments}
\label{Sec_Acknowledgements}

This work was carried out under approved Grant No. MSC/2020/000049 from the Science and Engineering Research Board, Government of India.

\enlargethispage{1.4cm}

\end{document}